\documentclass[conference,hidelinks]{IEEEtran}
\IEEEoverridecommandlockouts
\usepackage[utf8]{inputenc}

\usepackage[mathscr]{euscript} 
\usepackage[nolist]{acronym} 
\usepackage[normalem]{ulem} 
\usepackage[nospace,noadjust]{cite} 
\usepackage{algorithmic}
\usepackage{algorithm}
\usepackage{amsmath,amssymb,amsfonts} 
\usepackage{amsthm} 
\usepackage{array}
\usepackage{float}
\usepackage{flushend} 
\usepackage{gensymb}
\usepackage{mathrsfs} 
\usepackage{multirow} 
\usepackage{soul} 
\usepackage{textcomp} 
\usepackage{xcolor}

\usepackage{graphicx}
\usepackage[inkscapeformat=png]{svg}

\usepackage{tikz}
\tikzset{
  font={\fontsize{9pt}{12}\selectfont}}
\usetikzlibrary{plotmarks}
\usetikzlibrary{intersections,backgrounds, patterns}
\usetikzlibrary{patterns,shapes.arrows}

\usepackage{pgfplots}
\pgfplotsset{compat=newest} 
\usepgflibrary{arrows}
\usepgfplotslibrary{groupplots,dateplot}
\pgfplotsset{every axis/.append style={
	scaled x ticks = false,
	label style={font=\footnotesize},
	tick label style={font=\footnotesize},
	tick scale binop=\times}
}
\pgfkeys{/pgf/number format/.cd,
1000 sep={},
}

\usepackage[cmintegrals]{newtxmath} 
\usepackage[font=footnotesize]{subcaption}
\usepackage[font=footnotesize]{caption}
\usepackage{enumitem}
\DeclareUnicodeCharacter{2212}{−}

\definecolor{mycolor1}{RGB}{19, 133, 189}
\definecolor{mycolor2}{RGB}{230, 112, 32}
\definecolor{mycolor3}{RGB}{130, 173, 98}
\definecolor{mycolor4}{rgb}{0.49412,0.18431,0.55686}%
\definecolor{myhist}{RGB}{73, 80, 87}
\definecolor{hgreen}{rgb}{0, 0.5, 0}

\def\BibTeX{{\rm B\kern-.05em{\sc i\kern-.025em b}\kern-.08em
    T\kern-.1667em\lower.7ex\hbox{E}\kern-.125emX}}

\begin{document}

\title{Enabling Low-Overhead Over-the-Air Synchronization Using Online Learning\\
\thanks{The work of Hazem Sallouha was funded by the Research Foundation – Flanders (FWO), Postdoctoral Fellowship No. 12ZE222N.}
}


\author{\IEEEauthorblockN{Dieter Verbruggen, Hazem Sallouha, and Sofie Pollin}
\IEEEauthorblockA{\textit{Department of Electrical Engineering (ESAT) - WaveCoRE}\\
KU Leuven, 3000 Leuven, Belgium\\
E-mail:\{dieter.verbruggen, hazem.sallouha, sofie.pollin\}@kuleuven.be}}
\maketitle

\begin{abstract}
Accurate network synchronization is a key enabler for services such as coherent transmission, cooperative decoding, and localization in distributed and cell-free networks. 
Unlike centralized networks, where synchronization is generally needed between a user and a base station, synchronization in distributed networks needs to be maintained between several cooperative devices, which is an inherently challenging task due to hardware imperfections and environmental influences on the clock, such as temperature. As a result, distributed networks have to be frequently 
synchronized, introducing a 
significant synchronization overhead. In this paper, we propose an online-LSTM-based model for clock skew and drift compensation, to elongate the period at which synchronization signals are needed, 
decreasing the synchronization overhead. We conducted comprehensive experimental results to assess the performance of the proposed model. Our measurement-based results show that the proposed model reduces the need for re-synchronization between devices by an order of magnitude, keeping devices synchronized with a precision of at least 10 microseconds with a probability 90\%.
\end{abstract}


\section{Introduction}


Recent advances in wireless networks are showing a clear trend in moving away from the conventional centralized star-based architectures to distributed alternatives, addressing contention, interference, and coexistence challenges. This trend can be seen in most recent processing paradigms, such as edge-computing \cite{edge_comp}, and federated learning \cite{MAL-083} as well as networking paradigms, such as cell-free \cite{chen2021survey}, and crowdsourced networks \cite{rajendran2017electrosense}. In order to realize the full potential of these paradigms, accurate network synchronization is needed as a key enabler for essential services such as coherent transmission \cite{demir2021foundations}, cooperative decoding \cite{calvo2019collaborative}, and localization \cite{sallouha2021aerial}. While centralized networks can achieve high-precision synchronization via wired-based infrastructure or over-the-air (OTA) with acceptable overhead, distributed networks may not have wired-based infrastructure due to cost and geographical constraints \cite{sallouha2021aerial}, and existing OTA synchronization solutions are unscalable for distributed networks due to the excessive overhead needed to synchronize significantly more devices when compared to centralized networks \cite{interdonato2019ubiquitous,demir2021foundations,calvo2019collaborative,sallouha2021aerial,rajendran2017electrosense}.

The essence of network synchronization is to keep a target group of devices running with the same clock reference. However, due to imperfect hardware, represented by the clock circuits, as well as environmental factors, such as temperature, a device's local clock may deviate from the common reference. One solution is to have extra hardware with every device in the network, such as a GPS disciplined oscillator (GPSDO), acting as an accurate time source for individual devices \cite{sallouha2019localization}. However, extra-hardware-based solutions are typically expensive in terms of cost and power consumption. Another widely adopted synchronization method in modern wireless systems, such as 3GPP LTE (3rd Generation Partnership Project Long Term Evolution) standardization, relies on periodic synchronization pilots exchange between target devices and a reference access point \cite{conformance20113rd}. These pilot signals can be used in an oriented way, as in the case of LTE, or opportunistically in distributed networks, as in the case of crowdsourced networks \cite{calvo2019collaborative}. The tradeoff with this pilot-based synchronization method is the pilot overhead. For instance, in a distributed network, where multiple reference access points work cooperatively to serve multiple devices, the pilot-based synchronization overhead becomes overwhelming, significantly reducing the network's spectral efficiency \cite{interdonato2019ubiquitous}. In fact, the rate at which synchronization pilots are needed in distributed networks is determined by both the quality of the device's crystal oscillator and the application's maximum tolerated clock offset.

The model of digital clocks in wireless devices can be represented by a time-series process, i.e., discrete stochastic process \cite{masood2016disty}. An accurate clock model is a key enabler for clock drift prediction, which promises a reduced synchronization overhead by relying on the clock model to compensate for clock drifts, minimizing the frequency at which synchronization pilots are needed \cite{calvo2019collaborative,sallouha2021aerial}. Recent state-of-the-art works addressed the digital clock modeling by using autoregressive models along with a Kalman filter \cite{masood2016disty} or by exploiting long short-term memory (LSTM)-based recurrent neural networks (RNNs) \cite{sallouha2021aerial}. While both autoregressive and LSTM clock models showed promising performance in predicting the clock drifts, these models were trained in an offline manner, and hence frequent retraining would be needed to account for the time-varying nature of clock drifts as well as changes in the target environment \cite{kim2011tracking}. Online learning offers attractive solutions to address the need for comprehensive model retraining by exploiting the sequentially available training data to update and adapt the trained model \cite{hu2021distributed}. Online learning techniques have been employed in several areas of distributed wireless networks, such as inferring system conditions and performing adaptive resource allocations \cite{hu2021distributed}. The promising performance of online learning methods and the urgent need for synchronization methods with low pilot signals overhead and low training overhead drive our study in this paper.

\subsection{Related Works}

Several works in the literature presented methods to address the clock drift problem in large-scale networks, passively by using reference signals, such as LTE pilots or automatic dependent surveillance-broadcast (ADS-B) signals \cite{airsync}, or actively by propagating synchronization signals over the network \cite{PulseSync}. 
These state-of-the-art methods are mainly conducted for wireless sensor networks with limited cost and computational capabilities. While some works consider compensation methods based on linear programming and multi-cast \cite{linearprogramming}, most research reports are centered around temperature-assisted methods \cite{linearprogramming,Temp_assisted,Temperature}. This arises from the strong correlation between the working temperature and the oscillator frequency used in wireless devices. 

Obtaining an accurate temperature-frequency model requires continuous measurements of the oscillator frequency, which is a challenging task for devices with limited computational power. In \cite{Temp_assisted}, the authors considered a static model by which the temperature-frequency model is approximated using a second-order or third-order polynomial. The weights for the second-order and third-order polynomials are estimated using priory measurements. Haapala \textit{et al.} \cite{nB-iot} presented a dynamic model based on lookup tables and interpolation between known temperature-frequency pairs. However, these aforementioned works targeted low-end clock oscillators, which have a more outspoken temperature-frequency correlation compared to the high-end oscillators considered in this work. 

\subsection{Contribution and Paper Structure}
In this paper, we propose and experimentally evaluate a method to track the local clock of devices based on opportunistically existing signals, such as LTE signals, and environment-dependent features, such as temperature measurements. The proposed method exploits LTE signals opportunistically and combines them with an online LSTM-based prediction model, enabling accurate local clock drift prediction. This accurate prediction significantly reduces the need for re-synchronization between devices, and hence the overall network synchronization overhead. In particular, our measurement-based results show that the proposed model reduces the need for re-synchronization between devices from 2 min without any compensation to 55 minutes to keep devices synchronized with a precision of 10 microseconds in 90\% of the time.
The main contribution of this paper is twofold.
\begin{itemize}
\item First, we introduce a novel LSTM-based clock model that uses online learning, which unlike existing works in the literature, does not require comprehensive and frequent retraining. The proposed model predicts and compensates for devices' local clocks, minimizing the need for re-synchronization signals between devices. We rely on the fact that the majority of wireless transceivers use the same oscillator for both the local clock and the radio frequency (RF) front-end, which facilitates accurate measurements of the oscillator frequency by measuring the frequency offset of the RF front-end, enabling our proposed method to adapt the model based on the sequentially coming measurements in an online fashion.
\item Second, we conduct extensive measurements using off-the-shelf software-defined radios (SDR) to collect their local clock measurements, LTE pilot signals they receive, as well as temperature measurements. We consider SDRs as they can resemble interesting use cases such as cell-free and crowdsourced networks, and they can be used in many applications, some of which require strict time synchronization.
\end{itemize}

The rest of the paper is organized as follows. Section II presents the system model. In Section III, we introduce our proposed synchronization method. The experiment design and data collection are detailed in Section IV. Subsequently, we present the performance evaluation results in Section V. Finally, the paper is concluded in Section VI.

\section{System Model}


This paper considers a set of stationary transceivers, represented by SDRs, distributed randomly in a given area. The area of interest can be a mix of an indoor and outdoor environment. We assume that this target group of nodes is within communication range of an LTE base station, enabling us to use LTE signals opportunistically as a synchronization reference to measure the oscillator frequency. Our aim is to keep our target group of transceivers synchronized with the precision requirements for an application, such as collaborative spectrum sensing and cooperative decoding. 
Each node in the network has its own single oscillator (TG2016SBN)
to generate the local notion of time, represented by the clock ticks, for the RF front-end as well as for the baseband processing. 
In the literature, three different terms are commonly used when discussing the non-ideal behaviour of clocks: 
\begin{itemize}
    \item \textit{Offset} is the time difference between two clocks; when this value is zero, both clocks are perfectly synchronized.
    \item \textit{Skew} is the difference between clocks frequencies.
    \item \textit{Drift} refers to small variations on the skew, usually as a consequence of environmental changes such as temperature \cite{basics}.
\end{itemize} 
Throughout this paper, ppm (parts per million) is used as a way of generalizing the results as it is frequency-independent. Furthermore, using ppm enables the added benefit of quickly calculating the clock offset after a specific period. For example, an oscillator with a constant ppm of $0.1$ will result in a clock offset of $0.1\,\mu$sec after $1$\,sec. If this system requires a maximum clock offset of $6\,\mu$sec, the clocks must be synchronized each minute.

\section{LSTM-Based Online Learning Method}

In this section, we introduce our proposed LSTM-based clock model with its online training. Given the nature of the clock modeling problem, in which the true reference of the clock is measured or obtained via pilot signals sequentially, we adopted an online learning method to keep our LSTM-based model updated. This online method eliminates the need to retrain the underlying model with sizable training data, which translates into lower training overhead. We consider an RNN with one LSTM layer followed by a fully connected layer. The Adam optimizer \cite{adam} is used with a learning of 0.001 to train this network. Table \ref{tab:lstm_overview} details the network architecture.
The goal of this LSTM-based network is to estimate the ppm of the oscillator, exploiting the fact that varying temperature is one of the main causes for clock drifts \cite{sallouha2021aerial}.
In the following, we detail the different design parameters of our proposed approach.

\begin{table}[t]
    \caption{RNN Model architecture.}
    \label{tab:lstm_overview}
    \centering
    \begin{tabular}{ | c || c | c |}
    \hline
    Layer index& Type& Details \\\hline \hline
    1&Input&Units: $5$\\ \hline
    && Hidden States: $24$\\ \cline{3-3}
    2&LSTM& State activation: $tanh$\\ \cline{3-3}
    & & Gate activation: $sigmoid$\\ \hline
   3& Fully connected& Units: $24$\\\hline
\end{tabular}
\end{table}
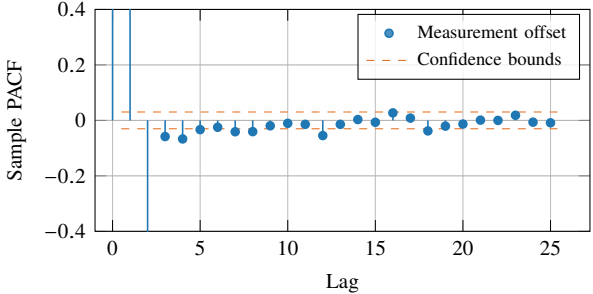
\begin{figure}[t]
\centering
\begin{tikzpicture}

\definecolor{crimson2143940}{RGB}{214,39,40}
\definecolor{darkgray176}{RGB}{176,176,176}
\definecolor{darkorange25512714}{RGB}{255,127,14}
\definecolor{forestgreen4416044}{RGB}{44,160,44}
\definecolor{lightgray204}{RGB}{204,204,204}
\definecolor{steelblue31119180}{RGB}{31,119,180}


\begin{axis}[
legend style={font=\scriptsize},
legend cell align={left},
legend style={fill opacity=0.8, draw opacity=1, text opacity=1},
height=0.25\textwidth,
width=0.45\textwidth,
xlabel={Lag},
ylabel={Sample PACF},
xmajorgrids,
ymajorgrids,
xmin=-1, xmax=27.25,
xtick style={color=black},
scaled x ticks=false,
xticklabel style={/pgf/number format/fixed,},
y grid style={darkgray176},
ymin=-0.4, ymax=0.4,
ytick={-0.2,-0.4,0,0.2,0.4}
]

\path [draw=steelblue31119180, semithick]
(axis cs:0,0)
--(axis cs:0,1);

\path [draw=steelblue31119180, semithick]
(axis cs:1,0)
--(axis cs:1,0.999585523463267);

\path [draw=steelblue31119180, semithick]
(axis cs:2,0)
--(axis cs:2,-0.449382100347911);

\path [draw=steelblue31119180, semithick]
(axis cs:3,0)
--(axis cs:3,-0.0582993438612476);

\path [draw=steelblue31119180, semithick]
(axis cs:4,0)
--(axis cs:4,-0.0668375223842156);

\path [draw=steelblue31119180, semithick]
(axis cs:5,0)
--(axis cs:5,-0.0332567181522147);

\path [draw=steelblue31119180, semithick]
(axis cs:6,0)
--(axis cs:6,-0.0248026135420589);

\path [draw=steelblue31119180, semithick]
(axis cs:7,0)
--(axis cs:7,-0.0409727969633158);

\path [draw=steelblue31119180, semithick]
(axis cs:8,0)
--(axis cs:8,-0.0406245631965078);

\path [draw=steelblue31119180, semithick]
(axis cs:9,0)
--(axis cs:9,-0.0195335667585805);

\path [draw=steelblue31119180, semithick]
(axis cs:10,0)
--(axis cs:10,-0.0106000192654195);

\path [draw=steelblue31119180, semithick]
(axis cs:11,0)
--(axis cs:11,-0.0142476425360449);

\path [draw=steelblue31119180, semithick]
(axis cs:12,0)
--(axis cs:12,-0.0547974872986944);

\path [draw=steelblue31119180, semithick]
(axis cs:13,0)
--(axis cs:13,-0.0143084826716854);

\path [draw=steelblue31119180, semithick]
(axis cs:14,0)
--(axis cs:14,0.0029931145237947);

\path [draw=steelblue31119180, semithick]
(axis cs:15,0)
--(axis cs:15,-0.00688914986575719);

\path [draw=steelblue31119180, semithick]
(axis cs:16,0)
--(axis cs:16,0.0270683886655651);

\path [draw=steelblue31119180, semithick]
(axis cs:17,0)
--(axis cs:17,0.00826135363368871);

\path [draw=steelblue31119180, semithick]
(axis cs:18,0)
--(axis cs:18,-0.0379577389186683);

\path [draw=steelblue31119180, semithick]
(axis cs:19,0)
--(axis cs:19,-0.0207452364910981);

\path [draw=steelblue31119180, semithick]
(axis cs:20,0)
--(axis cs:20,-0.0132508515303278);

\path [draw=steelblue31119180, semithick]
(axis cs:21,0)
--(axis cs:21,0.000876515303808182);

\path [draw=steelblue31119180, semithick]
(axis cs:22,0)
--(axis cs:22,-0.000108404757001296);

\path [draw=steelblue31119180, semithick]
(axis cs:23,0)
--(axis cs:23,0.0183826729630363);

\path [draw=steelblue31119180, semithick]
(axis cs:24,0)
--(axis cs:24,-0.00641734182665013);

\path [draw=steelblue31119180, semithick]
(axis cs:25,0)
--(axis cs:25,-0.0087187098415558);
\addplot [semithick, steelblue31119180, mark=*, mark size=1.5, mark options={solid}, only marks]
table {%
0 1
1 0.999585523463267
2 -0.449382100347911
3 -0.0582993438612476
4 -0.0668375223842156
5 -0.0332567181522147
6 -0.0248026135420589
7 -0.0409727969633158
8 -0.0406245631965078
9 -0.0195335667585805
10 -0.0106000192654195
11 -0.0142476425360449
12 -0.0547974872986944
13 -0.0143084826716854
14 0.0029931145237947
15 -0.00688914986575719
16 0.0270683886655651
17 0.00826135363368871
18 -0.0379577389186683
19 -0.0207452364910981
20 -0.0132508515303278
21 0.000876515303808182
22 -0.000108404757001296
23 0.0183826729630363
24 -0.00641734182665013
25 -0.0087187098415558
};
\addlegendentry{Measurement offset }

\addplot [dashed,mycolor2] coordinates {
(0.5,0.0302429008619579) (25.5,0.0302429008619579)
};
\addplot [dashed,mycolor2] coordinates {
(0.5,-0.0302429008619579) (25.5,-0.0302429008619579)
};
\addlegendentry{Confidence bounds }

\end{axis}

\end{tikzpicture}
\caption{The sample PACF of the measured clock offset with 95\% confidence interval.}
\label{fig:PACF}
\vspace{-0.3cm}
\end{figure}

\begin{enumerate}[leftmargin=*]
\item \textit{Lag}: The lag defines the number of samples from the past on which the current prediction depends. We rely on the PAFC, a known metric that indicates the impact of different lags, to select the lag. From Fig. \ref{fig:PACF}, it can be concluded that lags bigger than five do not have a significant impact as their correlations are below the 95\% confidence interval. Accordingly, a lag of five is chosen for the proposed model.   
\item \textit{Input features}: The input features considered in the proposed method are the timestamp and temperature. In particular, the timestamp is the seconds-of-the-day time counted from midnight.

\item \textit{State and gate activation functions}: For the state and gate activation functions, the default activation functions used in LSTM-based RNN are chosen, which are $tanh$ for the state activation and $sigmoid$ for the gate activation.

\item \textit{Number of initial epochs ($N_{\text{initial}}$) and hidden states}: The number of initial epochs and the hidden states are obtained using a Monte Carlo simulation, from which we selected the best-performing parameters. The number of initial epochs controls how often the model trains on the data. Values of 25 and 24 have been chosen empirically
for the number of initial epochs and the size of the hidden state, respectively, as they result in the best trade-off between learning the underlying model and learning the noise of the measurements.  
\item \textit{Time between online measurements} ($\Delta t_{\text{online}}$): We define the time $\Delta t_{\text{online}}$, as the period at which we update our LSTM-based model with new training data. For instance, a $\Delta t_{\text{online}} = 20$ min, means that we update our LSTM-based model with the new training data every 20 minutes.
\item \textit{Number of online epochs ($N_{\text{online}}$)}: The number of online epochs represents the number of times the model trains on the sequentially arriving data points. Intuitively, the number of online epochs depends on the chosen $\Delta t_{\text{online}}$. A lower $\Delta t_{\text{online}}$ requires a lower number of epochs as the RNN is retrained more frequently over shorter periods of time.
\end{enumerate}
Considering the design parameters of the proposed LSTM-based online learning method, in the following section, we design an experiment to assess the performance of the proposed method with real-life measurements.

\section{Experiment Design and Data Collection}

In this section, we introduce our experiment setup and the data collection process, in which we focus on the node level, aiming to compensate for any drift and skew resulting from the hardware imperfection. 

\subsection{Experiment Setup}
\begin{figure}
     \centering
    \includegraphics[width=0.9\linewidth]{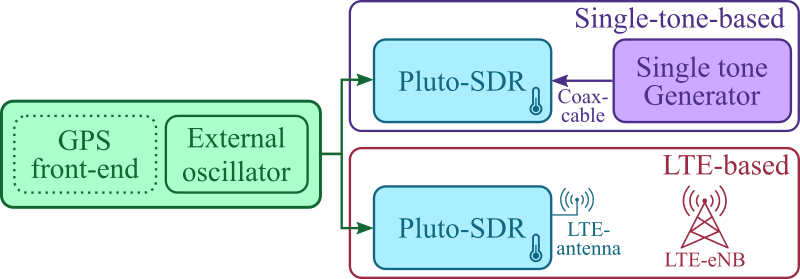}
    \caption{Experiment setup for Data collection (with/without GPS front-end).}
    \label{fig:experiment_setup}
 \vspace{-0.3cm}   
\end{figure}

The experiment setup depicted in Fig. \ref{fig:experiment_setup}, consists of two Pluto-SDRs with a common oscillator, reference clock, and a single-tone generator. 
   

 
\begin{itemize}
    \item The Pluto-SDR\footnote{https://wiki.analog.com/university/tools/pluto}, produced by Analog Devices, is a low-cost SDR used in academia and for educational purposes. The Pluto-SDR is widely adopted in the wireless community for a broad range of applications, and it has an advanced RF front-end, making it an appealing candidate for our experiments. However, this off-the-shelf SDR has a low-end oscillator with a quality of 25 ppm, which is unsuitable for time-sensitive and synchronization purposes. This low-end oscillator can be bypassed by connecting an external oscillator, easily replacing the low-end oscillator with a high-end one on-demand.
    \item The common external oscillator for this experiment is the Mini Precision GPS Reference Clock designed by Leo Bodnar\footnote{http://www.leobodnar.com}. This clock is programmable for almost all frequencies between 400Hz and 810MHz; some frequencies are not achievable due to hardware limitations. With the GPS antenna attached, this oscillator is GPS disciplined, approaching $1\mathrm{e}{-6}$ ppm. Without the GPS antenna, the stability of the oscillator is dictated by the high-end internal oscillator (TG2016SBN), making it prone to the environment with a ppm of 0.5 between -30 and 85 \celsius.  
    \item The single-tone generator is an extra Pluto-SDR with a GPSDO. This Pluto-SDR transmits a single sine wave of chosen frequency on a carrier frequency. Due to the stable oscillator, the generated signal and the carrier frequency are accurate and stable. 
\end{itemize}
As shown in Fig. \ref{fig:experiment_setup}, the two Pluto-SDRs share the same external oscillator and measure the ppm of this external oscillator with different methods; using single-tone as a reference for the first one, and LTE Primary Synchronization Signals (PSS) as a reference for the second one. Both these methods exploit the RF front-end of the Pluto-SDR to measure the oscillator's ppm. An offset between the frequency of the oscillator $f_{xo}$ and its nominal frequency $f_{xo,nom}$ will result in an offset between the sampling frequency $f_{s}$ and its nominal frequency $f_{s,nom}$. This relation between frequencies and the corresponding ppm can be described as
\begin{equation}
    \text{pmm}=\frac{f_{x}-f_{x,nom}}{f_{x,nom}}10^{6}\,, \quad \forall x \in \{ xo, s, sine\}\,,
    \label{eq:ppm}
\end{equation}
where subscripts $xo$, $s$, and $sine$ are used to refer to oscillator, sampling, and sine wave, respectively.
In the following, we detail the single-tone and the LTE as references for our measurements.
\subsubsection{Single-tone-based as a benchmark} The single-tone generator transmits a single sine wave with a frequency $f_{sine,nom}$ of 160\,kHz on a carrier frequency of 2.4\,GHz over a coaxial cable. When the receiver's sampling frequency $f_{s}$ deviates from the nominal sampling rate $f_{s,nom}$ of 5\,Msps, the received sine wave will be received with a deviated frequency $f_{sine}$. Accordingly, the ppm can be calculated using (\ref{eq:ppm}) with subscript $x$ being $sine$. It is worth noting that this single-tone method, commonly used for SDR's calibration, is only used as a benchmark reference for our LSTM-based method with LTE signals .

\subsubsection{LTE-based} Our LTE-based method relies on counting the samples in between PSS, which are transmitted every $5ms$ by Frequency Division Duplex (FDD)-LTE base stations \cite{Ltesstrack}. This method captures one second of samples containing around 200 PSS. The PSS signals are detected using time correlation, up-sampling, and peak detection. By calculating the average number of samples between peaks, the effective sampling frequency $f_{s}$ can be calculated, and thus the $f_{xo}$ and ppm, using (\ref{eq:ppm}). This LTE-based method is a promising solution for the synchronization problem in networks with large-scale deployments, such as crowdsourced networks \cite{rajendran2017electrosense}.

\subsection{Data Collection and Model Training}


\subsubsection{Data collection}
In order to collect the dataset for the performance evaluation of the proposed method, we used the setup shown in Fig. \ref{fig:experiment_setup} with the GPS front-end disconnected, representing a node in the system model without GPSDO. This experiment has been conducted outdoors to capture a wide range of temperatures compared to the slow-varying temperature indoors. The collected data consists of 1) timestamp (second of the day), 2) temperature (measurements collected by the onboard temperature sensor of the Pluto-SDR), 3) LTE-based ppm measurement, and 4) single-tone-based ppm measurement (benchmark). The measurements were collected continuously and averaged per one-minute window to improve the precision and reduce the noise of the measurements (cf. \ref{sec:max_precis}). In total, 4,200 data points were collected over a span of 70 hours.

\subsubsection{Training and prediction}
In order to train and assess our proposed online LSTM-based method, we used 1440 data points corresponding to the measurements taken on the first day as the initial training data for the proposed LSTM model. Subsequently, the remaining data is divided into two groups for our online learning method. The first group is used for testing our online prediction using temperature and timestamp as input features, and the second group is used sequentially for our proposed LSTM online learning, i.e., updating and adapting the model when new labelled data arrives.

\section{Performance Evaluation Results}
In this section, we assess the accuracy of the single-tone-based and LTE-based ppm measurements and assess the performance of the proposed approach. 





\subsection{Ppm Measurement Accuracy} 
\label{sec:max_precis}
To assess the accuracy of the single-tone-based and LTE-based ppm measurements, the experiment setup shown by Fig. \ref{fig:experiment_setup} is used. We connected the GPS front-end, improving the external oscillator's stability. The external oscillator is programmed to sweep from $-0.5$ ppm till $0.5$ ppm with a step of $0.025$ ppm, simulating an oscillator with a fixed skew. We measure the  oscillator ppm using the single-tone-based and LTE-based methods and compare the measurements against the corresponding programmed ppm. The resulting measurements are filtered for outliers, and some steps in the sweep have been omitted as empirical comparison concluded that the external oscillator could not accurately produce this ppm. In total, we collected around 21,400 single-tone-based and 2,840 LTE-based measurements. 

\begin{table}[t]
    \caption{Details of the ppm measurements.}
    \label{tab:accuracy}
    \centering
    \begin{tabular}{ | c | c | c |}
    \cline{2-3}
    \multicolumn{1}{c|}{}  & \multicolumn{1}{c|}{Single-tone-based} & \multicolumn{1}{c|}{LTE-based} \\
    \hline
    Accuracy & $0.395\mathrm{e}{-3}$ &$0.812\mathrm{e}{-3}$  \\ 
    \hline
    Precision& $0.396\mathrm{e}{-3}$ &$16.602\mathrm{e}{-3}$\\ 
    \hline
\end{tabular}
\vspace{-0.3cm}
\end{table}

Table \ref{tab:accuracy} shows the details of the ppm measurements. The accuracy, in Table \ref{tab:accuracy}, indicates the bias in the measurements and is defined as the average difference between the measurement and the corresponding programmed ppm. Both methods show a bias in their measurements. However, in both cases, the bias is relatively small, e.g., a bias of $1\mathrm{e}{-3}$ introduces a clock offset of 1 nanosec every second or 3.6\,$\micro$sec every hour.  
The precision of the measurements, shown in Table \ref{tab:accuracy}, is defined as the standard deviation of the difference between the measurement and the corresponding programmed ppm. The measurements of the single-tone-based method are precise, introducing less than $25ns$ of error per minute. Confirming that the single-tone-based measurements can be used as a benchmark for our model. The LTE-based method is less precise than the single-tone-based method, as it introduces an error 1\,$\micro$sec each minute. The precision can be further improved by averaging over multiple measurements. The averaging introduces a trade-off between measurement accuracy and measurement time. In our data collection, averaging over one minute is chosen.


\begin{figure}[t]
     \centering
     \begin{subfigure}[b]{0.24\textwidth}
        \centering
\begin{tikzpicture}

\definecolor{crimson2143940}{RGB}{214,39,40}
\definecolor{darkgray176}{RGB}{176,176,176}
\definecolor{darkorange25512714}{RGB}{255,127,14}
\definecolor{forestgreen4416044}{RGB}{44,160,44}
\definecolor{lightgray204}{RGB}{204,204,204}
\definecolor{steelblue31119180}{RGB}{31,119,180}
\begin{axis}[
legend columns=-1,
legend entries={Single-tone-based, LTE-based},
legend cell align={left},
legend style={font=\scriptsize},
legend style={fill opacity=0.8, draw opacity=1, text opacity=1},
legend style={at={(axis cs:0.005,1.03)},anchor=south west},
xlabel={Residual error in ppm},
ylabel={Prob. Density},
xmajorgrids,
ymajorgrids,
xmin=-0.05, xmax=0.05,
xticklabel style={/pgf/number format/fixed,font=\tiny},
yticklabel style={/pgf/number format/fixed,font=\scriptsize},
ymin=-0.0, ymax=1,
width=\textwidth,
]
\addplot [semithick, mark=triangle, crimson2143940,mark size=1.5pt]
table {%
-0.000165785323828459 0.00235017626321974
-0.00014034479111433 0.00235017626321974
-0.000114904258400202 0.000783392087739914
-8.94637256860733e-05 0.0176263219741481
-6.40231929719448e-05 0.00313356835095966
-3.85826602578163e-05 0.00783392087739914
-1.31421275436879e-05 0.0470035252643948
1.22984051704407e-05 0
3.77389378845692e-05 0.0771641206423815
6.31794705986977e-05 0.104191147669409
8.86200033128261e-05 0.0955738347042695
0.000114060536026955 0.225616921269095
0.000139501068741083 0.072855464159812
0.000164941601455212 0.374461417939679
0.00019038213416934 0
0.000215822666883469 0.244026635330983
0.000241263199597597 0.304739522130827
0.000266703732311726 0.150019584802193
0.000292144265025854 0.715236976106541
0.000317584797739983 0.169996083039561
0.000343025330454111 0.389345867606737
0.00036846586316824 0.667450058754407
0.000393906395882368 0.413631022326674
0.000419346928596497 0.426165295730513
0.000444787461310625 0.200548374461418
0.000470227994024754 1
0.000495668526738882 0.387387387387387
0.000521109059453011 0.535056795926361
0.000546549592167139 0.319232275754015
0.000571990124881268 0.286329808068938
0.000597430657595396 0.245593419506463
0.000622871190309525 0
0.000648311723023653 0.280454367410889
0.000673752255737782 0.063454759106933
0.00069919278845191 0.267920094007051
0.000724633321166039 0.0466118292205249
0.000750073853880167 0.045436741088915
0.000775514386594295 0.105366235801018
0.000800954919308424 0
0.000826395452022553 0.0356443399921661
0.000851835984736681 0.0160595377986682
0.000877276517450809 0.0168429298864081
0.000902717050164938 0.00626713670191931
0.000928157582879066 0.00665883274578927
0.000953598115593195 0.00470035252643948
0.000979038648307323 0
0.00100447918102145 0.000391696043869957
0.00102991971373558 0
0.00105536024644971 0
0.00108080077916384 0.00117508813160987
};

\addplot [semithick, mark=o, darkorange25512714,mark size=1.5pt]
table {%
-0.0446755703799427 0.0333333333333333
-0.0426692434176803 0.0222222222222222
-0.0406629164554179 0.0555555555555556
-0.0386565894931555 0.0555555555555556
-0.0366502625308931 0.0555555555555556
-0.0346439355686307 0.1
-0.0326376086063683 0.116666666666667
-0.0306312816441059 0.177777777777778
-0.0286249546818435 0.133333333333333
-0.0266186277195811 0.205555555555556
-0.0246123007573187 0.216666666666667
-0.0226059737950563 0.277777777777778
-0.020599646832794 0.255555555555556
-0.0185933198705316 0.277777777777778
-0.0165869929082692 0.283333333333333
-0.0145806659460068 0.272222222222222
-0.0125743389837444 0.355555555555556
-0.010568012021482 0.483333333333333
-0.0085616850592196 0.6
-0.00655535809695721 0.605555555555556
-0.00454903113469482 0.866666666666667
-0.00254270417243243 0.872222222222222
-0.000536377210170039 0.872222222222222
0.00146994975209235 0.938888888888889
0.00347627671435475 1
0.00548260367661714 0.844444444444444
0.00748893063887953 0.811111111111111
0.00949525760114192 0.75
0.0115015845634043 0.622222222222222
0.0135079115256667 0.477777777777778
0.0155142384879291 0.522222222222222
0.0175205654501915 0.322222222222222
0.0195268924124539 0.344444444444444
0.0215332193747163 0.277777777777778
0.0235395463369787 0.194444444444444
0.0255458732992411 0.266666666666667
0.0275522002615034 0.25
0.0295585272237658 0.166666666666667
0.0315648541860282 0.161111111111111
0.0335711811482906 0.161111111111111
0.035577508110553 0.111111111111111
0.0375838350728154 0.0555555555555556
0.0395901620350778 0.0722222222222222
0.0415964889973402 0.0555555555555556
0.0436028159596026 0.0611111111111111
0.045609142921865 0.05
0.0476154698841274 0.0166666666666667
0.0496217968463898 0.0111111111111111
0.0516281238086522 0.0166666666666667
0.0536344507709146 0.0166666666666667
};

\end{axis}

\end{tikzpicture}
        \vspace{-0.5cm}
        \caption{}
        \label{fig:residual}
     \end{subfigure}
     \hfill
     \begin{subfigure}[b]{0.24\textwidth}
        \centering
\begin{tikzpicture}

\definecolor{darkgray176}{RGB}{176,176,176}
\definecolor{darkorange25512714}{RGB}{255,127,14}
\definecolor{lightgray204}{RGB}{204,204,204}
\definecolor{steelblue31119180}{RGB}{31,119,180}

\definecolor{crimson2143940}{RGB}{214,39,40}
\begin{axis}[
xlabel={Oscillator's ppm},
xmin=-0.4, xmax=0.4,
ylabel={Precision},
xticklabel style={/pgf/number format/fixed,font=\tiny},
scaled y ticks=false,
yticklabel style={/pgf/number format/fixed,font=\scriptsize},
ymin=-0.0, ymax=0.039,
xmajorgrids,
ymajorgrids,
width=\textwidth,
]
\addplot [semithick, mark=triangle, crimson2143940, mark size=1.5pt]
table {%
-0.45 0.000141953328435715
-0.4 0.000164469198870169
-0.35 0.000190319145165077
-0.325 0.000119912967406122
-0.275 0.000121372248054756
-0.25 0.000191980397330861
-0.225 0.000139109199483421
-0.2 0.0001746527947558
-0.175 0.000167975800338555
-0.15 0.000163050394822751
-0.125 0.000188186172702582
-0.1 0.000126450082626002
-0.075 0.000196996834458587
-0.025 0.000193154822558024
0 0.000172026012854795
0.025 0.000186001758934706
0.1 0.000197705087557996
0.125 0.000152880390413743
0.175 0.000162000009666845
0.2 0.000195336063003553
0.225 0.000177272326278085
0.25 0.000180629364827407
0.275 0.000180455439200951
0.3 0.00017295943225009
0.325 0.000200463262314042
0.35 0.00016127878394163
0.375 0.000205482463074285
0.4 0.000168504793043499
0.45 0.000177183026009811
0.475 0.000199430143408332
0.5 0.000197130491217136

};
\addplot [semithick, mark=o, darkorange25512714,mark size=1.5pt]
table {%
-0.45 0.0144077999188536
-0.4 0.0108395098990086
-0.35 0.00565738670860198
-0.325 0.00574028826652109
-0.275 0.0129021151886158
-0.25 0.0185502741308161
-0.225 0.0203818700000204
-0.2 0.021590708963811
-0.175 0.023121793227295
-0.15 0.0241233414182909
-0.125 0.0257440909229045
-0.1 0.0239659751496454
-0.075 0.0185500456150566
-0.025 0.00667739433321164
0 0.00452786174311824
0.025 0.00871414857505053
0.1 0.0253356716963443
0.125 0.0310697208869093
0.175 0.0310580961057673
0.2 0.0254575091283995
0.225 0.020751255599062
0.25 0.0161570558333805
0.275 0.0107977686319741
0.3 0.00874573581462188
0.325 0.00527871164769076
0.35 0.00515932264956612
0.375 0.00573848708192192
0.4 0.00851499966584856
0.45 0.010038476247736
0.475 0.0091819571734861
0.5 0.00708261482659648
};
\legend{}
\end{axis}

\end{tikzpicture}
        \vspace{-0.5cm}
        \caption{}
        \label{fig:std_ppm}
     \end{subfigure}
        \caption{(a) Residual error distribution. (b) Precision per ppm for both LTE and single-tone.}
        \label{fig:details}
\vspace{-0.5cm}
\end{figure}
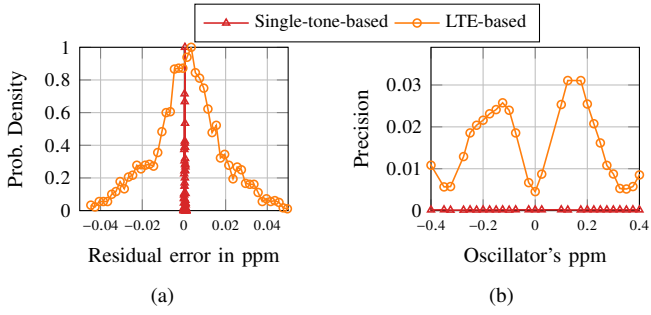
Fig. \ref{fig:residual} and Fig. \ref{fig:std_ppm} show, respectively, the residual error and the precision of the sweep with steps of 0.025 ppm. The residual error is defined as the difference between the programmed and measured ppm. The precision of the single-tone-based method is relatively constant compared to the almost LTE-based method. The precision of the LTE-based method increases in the regions around 0, -0.35, and 0.35 ppm as these values correspond to an integer amount of samples between PSS peaks. The decrease in the other regions is due to the limited accuracy of the peak detection.
\vspace{-0.1cm}

\subsection{Evaluation of the Proposed LSTM Model}
\begin{figure}[t]
    \centering
    \input{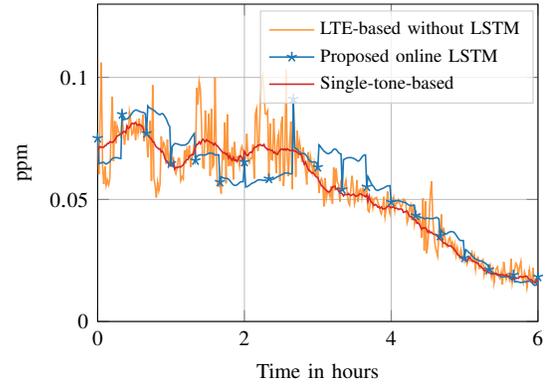}
    \caption{Samples of collected data using the single-tone-based method, LTE-based method and the predictions of the proposed online LSTM model. The stars show the measurements used for online learning.}
    \label{fig:present_data}
    \vspace{-0.1cm}
\end{figure}

In the previous subsection, we defined the maximum measurement accuracy of both the single-tone-based and LTE-based methods. In this subsection, we introduce the performance evaluation of our proposed model. Fig. \ref{fig:present_data} shows a sample of the ppm predicted by the proposed LSTM model, the LTE-based ppm measurements and the corresponding single-tone-based ppm measurements. Here the $\Delta t_{\text{online}}$ is chosen to be 20 minutes. The data points used for online learning are shown using stars. 

The ppm prediction of the proposed LTE-based LSTM model is compared against three compensation methods. The first compensation method, named \textit{No compensation}, does not compensate for the oscillator's ppm; the predicted ppm of this method is zero. \textit{Single-tone-based compensation} compensates the oscillator's ppm using a predicted ppm measured with the single-tone-based method; the predicted ppm of this method stays constant until the subsequent measurement. Similar to single-tone-based compensation, \textit{LTE-based without LSTM compensation} compensates with a constant predicted ppm, which is measured based on LTE without relying on any clock modeling. The predicted ppm of the proposed LSTM model is obtained by training, and subsequently adapting the model constantly in an online manner, with the corresponding temperatures and timestamps. 

The local notion of time is compensated using the predicted ppm. The smaller the difference between the predicted ppm and the actual ppm, the smaller the clock offset, which is defined as the difference between the compensated local notion of time and the global notion of time (i.e., the absolute time). We use this clock offset as the performance metric to compare the different compensation methods. Setting this clock offset to zero corresponds to perfect synchronization. We synchronize the local clock every $\Delta t_{\text{online}}$ minutes, combining the perfect synchronization with a ppm measurement of the oscillator and, in the case of the proposed online LSTM model, also with an update of the model.
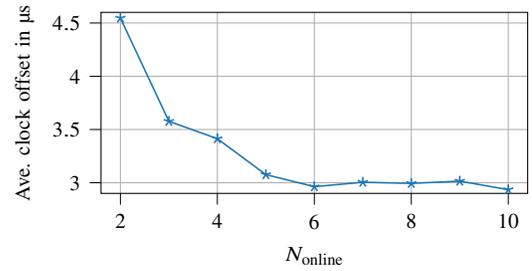
\begin{figure}[t]
    \centering
    \begin{tikzpicture}

\definecolor{darkgray176}{RGB}{176,176,176}
\definecolor{steelblue31119180}{RGB}{31,119,180}

\begin{axis}[
tick align=outside,
tick pos=left,
x grid style={darkgray176},
xlabel={$N_{\text{online}}$},
xmajorgrids,
xmin=1.6, xmax=10.4,
xtick style={color=black},
y grid style={darkgray176},
ylabel={Ave. clock offset in $\micro$s},
ymajorgrids,
ymin=2.90, ymax=4.6,
ytick style={color=black},
width=0.40\textwidth,
height=0.22\textwidth,
]
\addplot [semithick,mark=star, steelblue31119180]
table {%
2 4.54671415592746
3 3.57695800751906
4 3.41351970480823
5 3.07623999712365
6 2.96298587846001
7 3.00458414873123
8 2.99419813560517
9 3.0149703349732
10 2.93499276874446
};
\end{axis}

\end{tikzpicture}
    \caption{Average clock offset with $\Delta t_{\text{online}}$ of 25 minutes for the proposed online LSTM with different $N_{\text{online}}$.}
    \vspace{-0.5cm}
    \label{fig:compare_N}
\end{figure}

\begin{figure}[t]
    \centering
    \input{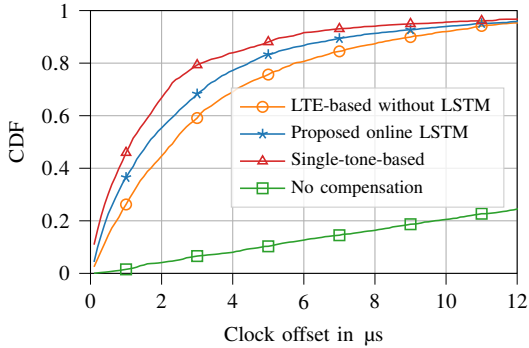}
    \caption{CDF of clock offset  with $\Delta t_{\text{online}}$ of 25 minutes with and without compensation.}
    \label{fig:cdf}
\end{figure}

The change in clock offset in microseconds after one minute is calculated by multiplying the difference between the actual ppm and the predicted ppm by 60. The clock offset at a specific moment can be calculated by adding all the previous changes in the clock offset starting from the previous perfect synchronization, as a perfect synchronization resets the clock offset. Fig. \ref{fig:compare_N} presents the average clock offset for the proposed online LSTM model with different $N_{\text{online}}$ considering $\Delta t_{\text{online}}$ of 25 minutes. Based on the figure, we opt for $N_{\text{online}}$ of six, since increasing $N_{\text{online}}$ further does not provide noticeable performance gains for the corresponding $\Delta t_{\text{online}}$ of 25 minutes. Fig. \ref{fig:cdf} shows the cumulative distribution function (CDF) of the clock offset for different compensation methods by which the system is synchronized every 25 minutes.  A lower clock offset means a better synchronization performance. Among these four compensation methods, No compensation performs the worse. Compensating the ppm is necessary to enable low-overhead synchronization. Fig. \ref{fig:cdf} shows that our proposed online LSTM-based method, with $N_{\text{online}}$ equals to six, outperforms the LTE-based compensation without LSTM, closing the gap compared to the single-tone-based compensation benchmark.


Table \ref{tab:time} presents the required $\Delta t_{\text{online}}$ for the considered compensation methods. An optimal $\Delta t_{\text{online}}$ can be chosen depending on the network's synchronization requirements and the maximum allowed synchronization overhead. For example, when a system requires a maximum clock offset of 10\,$\micro$sec for 90\% of the time, the corresponding $\Delta t_{\text{online}}$ can be measured for the different compensation methods considered as shown in Table \ref{tab:time}. Considering the proposed method, the network only has to re-synchronize every 55 minutes. Moreover, Table \ref{tab:time} illustrates that the proposed method reduces the synchronization overhead by around 95\% compared to \textit{No compensation} and more than 50\% compared to the \textit{LTE-based without LSTM} compensation. Finally, the computation time of our proposed method is assessed on a Raspberry Pi 4.
The proposed method's computation times for initial training, prediction, and online learning phases are 14.9 seconds, 22.52 ms, and 242.1 ms, respectively.

\begin{table}[t]
    \caption{$\Delta t_{\text{online}}$ for a clock offset less than 10\,$\mu$sec for 90\% of the time.}
    \label{tab:time}
    \centering
    \begin{tabular}{ | l | c | }
    \hline
    Method & $\Delta t_{\text{online}}$ \\
    \hline
    \hline
    No compensations & 2 minutes  \\ 
    \hline
        LTE-based without LSTM & 26 minutes  \\ 
    \hline
        Single-tone-based & 55 minutes  \\ 
    \hline
        Proposed online LSTM & 55 minutes  \\ 
    \hline
\end{tabular}
\vspace{-0.5cm} 
\end{table}




\section{Conclusion}
We proposed an online LSTM model for clock skew and drift compensation by exploiting temperature and LTE-based oscillator's ppm measurements, aiming to reduce over-the-air synchronization overhead in distributed wireless networks. The proposed model has been validated with real-life measurements and compared to different compensation methods, such as constant compensation with LTE-based and single-tone-based measurements. Our results showed that with LTE-based ppm measurements, the proposed online LSTM method could reduce the synchronization overhead needed to maintain a synchronization precision of 10\,$\micro$sec by several tens of minutes compared to methods with periodic synchronization signals without online learning. Fine-tuning the proposed method to be more application specific is an interesting future work.
\vspace{-0.2cm}



\bibliographystyle{ieeetr}
\bibliography{bibliography}

\end{document}